\documentclass[runningheads]{llncs}
\usepackage[T1]{fontenc}
\usepackage{graphicx}
\usepackage{comment}
\usepackage{cite}
\usepackage{multirow}
\usepackage{amsmath,amssymb,amsfonts}
\usepackage{algorithmic}
\usepackage{graphicx}
\usepackage{textcomp}
\usepackage{xcolor}
\usepackage{todonotes}
\usepackage{subcaption}
\usepackage{caption}
\usepackage{todonotes}
\usepackage[linesnumbered, ruled,vlined]{algorithm2e}
\usepackage{lipsum}
\usepackage{url}
\usepackage{authblk}
\usepackage{blindtext}
\usepackage[switch]{lineno}
\usepackage{booktabs}
\usepackage{mathrsfs} % 花体
\usepackage{listings}
\usepackage{float}
\usepackage{enumitem}
\usepackage{balance}
\usepackage{flushend}

\usepackage{hyperref} % Add this line to your preamble

\SetKwFor{For}{for}{}{}
\SetKwRepeat{Repeat}{repeat}{until}
\SetKwIF{If}{ElseIf}{Else}{if}{}{else if}{else}{}

\usepackage{xspace}
\usepackage[nolist]{acronym}
\usepackage{enumerate}
\usepackage{enumitem}
\usepackage{ifthen}

\newboolean{GuideOn} %control to hide GB guidelines
\setboolean{GuideOn}{false}

\newcommand{\ourname}{{GEMM-MP}\xspace}

\newcommand{\fugaku}{{Fugaku}\xspace}
\newcommand{\frontier}{{Frontier}\xspace}

\newcommand{\guyot}{{Guyot}\xspace}

\newcommand{\parsec}{{PaRSEC}\xspace}

\newcommand{\reduce}[1]{}

\begin{document}
%
%\title{Scalable and Efficient Mixed-Precision \\General Matrix Multiply}
\title{Leveraging Hardware-Aware Computation in Mixed-Precision Matrix Multiply: \\A Tile-Centric Approach}
\titlerunning{Mixed-Precision GEMM}
\author{
Qiao Zhang\inst{1} \and
Rabab Alomairy\inst{2, 3} \and 
Dali Wang\inst{4} \and
Zhuowei Gu\inst{1} \and
Qinglei Cao\inst{1}
}
%rabab.alomairy@mit.edu
%
\authorrunning{Qiao Zhang et al.}
\institute{
Saint Louis University, USA \\ \and
Massachusetts Institute of Technology, USA \\ \and
King Abdullah University of Science and Technology, KSA \\ \and
Oak Ridge National Laboratory, USA
}
\begin{comment}
\author{
Qiao Zhang\inst{1} \and
Rabab Alomairy\inst{2, 3} \and 
Dali Wang\inst{4} \and
Zhuowei Gu\inst{1} \and
Qinglei Cao\inst{1}
}
%rabab.alomairy@mit.edu
%
\authorrunning{Qiao Zhang et al.}
\institute{
Saint Louis University, USA \\ \and
Massachusetts Institute of Technology, USA \\ \and
King Abdullah University of Science and Technology, KSA \\ \and
Oak Ridge National Laboratory, USA
}
\end{comment}
%
\maketitle              % typeset the header of the contribution
\begin{abstract}
%The abstract should briefly summarize the contents of the paper in 150--250 words.

General Matrix Multiplication (GEMM) is a critical operation underpinning a wide range of applications in high-performance computing (HPC) and artificial intelligence (AI). The emergence of hardware optimized for low-precision arithmetic necessitates a reevaluation of numerical algorithms to leverage mixed-precision computations, achieving improved performance and energy efficiency. 
This research introduces an adaptive mixed-precision GEMM framework that supports different precision formats at fine-grained tile/block levels. We utilize the \parsec runtime system to balance workloads across various architectures. The performance scales well on ARM CPU-based \fugaku supercomputer, Nvidia GPU-based A100 DGX, and AMD GPU-based \frontier supercomputer. This research aims to enhance computational efficiency and accuracy by bridging algorithmic advancements and hardware innovations, driving transformative progress in various applications.

\end{abstract}
%
%
%
%SP + DP is enough for WAMTA

%CPU (Fugaku) + GPU (ICL V100, A100, and H100 + Frontier)

\section{Introduction}

%core of linear algebra~\cite{dongarra1990set,abdelfattah2019towards} 

General Matrix Multiplication (GEMM) lies at the core of linear algebra and acts as a critical kernel in numerous High-Performance Computing (HPC) applications---such as earthquake simulation~\cite{ichimura2018fast} and weather and climate prediction~\cite{abdulah2024boosting}---and in Artificial Intelligence (AI) applications, including fully connected layers in traditional neural networks and natural language processing~\cite{vaswani2017attention}. GEMM is a computationally expensive operation due to its high demands on both memory and computation, whose performance directly affects the efficiency of these applications, making it a focal point for optimization on diverse architectures, ranging from homogeneous CPU clusters to heterogeneous GPU systems.

Over the last ten years, hardware designers have prioritized crafting processors that excel in high-speed, energy-efficient, and reduced-precision computations~\cite{top500}. A notable example is the widespread use of Nvidia GPUs in both HPC and AI tasks, driven by their exceptional processing capabilities and efficiency. The latest Nvidia GPUs, including the H100 and the upcoming B100, incorporate a range of precision formats, each influencing peak theoretical throughput differently. The lowest precision levels (FP8/INT8 on H100 and FP4 on B100) achieve speedups of $58 \times$ and $233 \times$, respectively, relative to FP64 computations, underscoring the dramatic acceleration possible when shifting from conventional double-precision arithmetic to Tensor Core-accelerated lower-precision execution. A similar evolution is observed in CPUs, where FP64 performance is typically double that of FP32 and quadruple that of FP16. This transition has motivated researchers to reassess classical numerical techniques, identifying areas where reduced precision is sufficient without compromising the overall quality of solutions. Additionally, leveraging lower precision not only diminishes memory demands—allowing more data to fit within high-speed memory and thereby improving execution rates~\cite{abdulah2021accelerating, cao2023reducing}—but also cuts down on energy consumption, making it a compelling strategy for boosting energy efficiency in computational workflows~\cite{haidar2018design}.

The synergy between hardware advancements and algorithmic refinements plays a pivotal role in tackling complex computational challenges. Mixed-precision computations, which involve utilizing multiple numerical precisions within a single task, have emerged as a powerful strategy~\cite{haidar2020mixed,netti2023mixed}. This methodology has been successfully integrated into numerous scientific and engineering disciplines, such as deep learning~\cite{nandakumar2020mixed}, computational fluid dynamics~\cite{walden2019mixed,brogi2024floating}, astronomy~\cite{doucet2019mixed}, climate modeling~\cite{chen2024mixed}, molecular dynamics~\cite{jia2020pushing}, and quantum mechanics~\cite{boku2017mixed}. The primary advantage of these techniques is their ability to harness lower-precision hardware capabilities to enhance computational speed while preserving necessary accuracy constraints. Nevertheless, most existing implementations rely on coarse-grained precision adjustments and are optimized for single-GPU or single-node setups, which limits their applicability to large-scale distributed environments. Efficiently handling mixed-precision computations in distributed multi-GPU architectures introduces significant hurdles due to disparities in data representation and precision consistency. Moreover, many existing solutions are tailored to specific hardware platforms, reducing their portability across diverse computing architectures.

This paper introduces a mixed-precision GEMM framework (\ourname) at fine-grained tile/block levels and integrates \ourname into the runtime system \parsec. The contributions of this paper are as follows:
\begin{itemize} [noitemsep, topsep=0pt]
    \item Introducing a tile-centric mixed-precision GEMM algorithm that carefully balances accuracy and performance, enabling a robust and trustworthy mixed-precision solution for HPC and AI applications.
    \item Leveraging a dynamic task-based runtime system to efficiently orchestrate heterogeneous tasks, dynamically adapting to workload variations while minimizing scheduling overhead and maximizing resource utilization.
    \item Demonstrating strong scalability and portability across a diverse range of homogeneous and heterogeneous computing platforms, ensuring broad applicability and efficiency.
\end{itemize}
To the best of our knowledge, this is the first work to introduce a tile-centric mixed-precision GEMM algorithm.

The structure of this paper is as follows. Section~\ref{sec:rw} discusses prior research relevant to our work. In Section~\ref{sec:bg}, we provide an overview of \parsec. The tile-centric mixed-precision GEMM framework is described in Section~\ref{sec:algo}. A performance evaluation is then conducted in Section~\ref{sec:perf}, and finally, Section~\ref{sec:conclusion} outlines our conclusions and future directions.

\section{Related Work}
\label{sec:rw}

\subsection{Mixed-Precision Matrix Computations}

% reduce the number of references
%Mixed-precision approaches have garnered extensive interest~\cite{blanchard2020mixed,sun2022dissecting,haidar2020mixed,abdelfattah2020investigating,netti2023mixed} and are widely applied across multiple computational fields~\cite{gong2019mixed,nandakumar2020mixed,walden2019mixed,brogi2024floating,seznec2018study,doucet2019mixed,paxton2022climate,chen2024mixed,le2013spfp,jia2020pushing,olivares2010accelerating,boku2017mixed}. A well-known example is iterative refinement techniques for linear solvers, where the system is initially solved using lower precision, followed by higher precision corrections applied iteratively to refine the residual. This strategy has demonstrated significant reductions in computational time and energy consumption, making it an effective approach for enhancing efficiency in large-scale computations~\cite{buttari2007mixed,haidar2018design,baboulin2009accelerating,haidar2018harnessing}. However, the effectiveness of such methods is highly dependent on the matrix condition number, and the requirement to store multiple precision copies of matrices can introduce substantial memory overhead.

Mixed-precision approaches have garnered extensive interest~\cite{haidar2020mixed,netti2023mixed} and are widely applied across multiple computational fields~\cite{nandakumar2020mixed,walden2019mixed,brogi2024floating,doucet2019mixed,chen2024mixed,jia2020pushing,boku2017mixed}. A well-known example is iterative refinement techniques for linear solvers, where the system is initially solved using lower precision, followed by higher precision corrections applied iteratively to refine the residual. This strategy has demonstrated significant reductions in computational time and energy consumption, making it an effective approach for enhancing efficiency in large-scale computations~\cite{haidar2018design,haidar2018harnessing}. However, the effectiveness of such methods is highly dependent on the matrix condition number, and the requirement to store multiple precision copies of matrices can introduce substantial memory overhead.

Ongoing research explores performance optimization while ensuring numerical reliability by selectively employing reduced precision in non-critical computations while preserving higher precision where necessary~\cite{cao2022reshaping,abdulah2024boosting,hatem2024toward}. These methodologies primarily target linear system solutions. In deep learning, reduced precision is typically applied to forward and backward computations, whereas weight updates often require higher precision accumulation to maintain numerical stability. This trade-off, however, may lead to accuracy degradation due to precision loss in key computations of forward and backward propagation. 
In this work, we introduce an adaptive tile-centric mixed-precision GEMM framework that dynamically adjusts precision levels within a single computation. This fine-grained approach can provide another solution to ensure that accuracy remains uncompromised by recognizing that not all computational components require the same precision level in HPC and applications.

%Despite these advancements, several important questions remain regarding the practical adoption of mixed-precision techniques. First, it is unclear how many applications currently support mixed-precision at a production level and how many workloads remain inherently constrained to FP64 due to intrinsic factors. Second, the impact of mixed-precision on the nature and severity of failures in HPC and AI applications, compared to traditional FP64, remains an open question. This uncertainty arises from the significant differences in numerical accuracy and range between lower precisions and FP64, forcing HPC and AI applications to target varying output accuracy requirements based on the employed numerical strategy~\cite{bauer2012performance}. In this paper, we introduce a fine-grained mixed-precision GEMM framework, providing another solution which may not affect the overall accuracy, because not every part of the computation demands the same precision level in many scientific and AI applications.

\subsection{Runtime Systems}

Recent advancements in task-based programming models have driven the scalability of scientific applications on modern supercomputers. A variety of runtime systems have been developed to facilitate efficient task execution, optimizing resource utilization across diverse architectures.
{OpenMP}~\cite{OpenMP} has emerged as a widely adopted standard for shared-memory parallelism, incorporating tasking constructs that provide dynamic scheduling and fine-grained control. By leveraging compiler directives and runtime support, OpenMP simplifies concurrency management and enhances performance in multi-threaded environments.
Building upon OpenMP, {OmpSs}~\cite{ompss} extends capabilities for heterogeneous architectures, introducing advanced dependency tracking and asynchronous execution strategies. Similarly, {COMPSs}~\cite{lordan2014servicess} provides a comprehensive programming environment, streamlining task parallelism across distributed computing infrastructures.
For heterogeneous, distributed computing platforms, {StarPU}~\cite{starpu} delivers a flexible runtime system that integrates scheduling mechanisms with explicit kernel annotations, optimizing execution across CPUs and GPUs. Meanwhile, {HPX}~\cite{Heller2013a} leverages the ParalleX model to support highly scalable applications through lightweight threading and asynchronous execution paradigms in modern C++.
Another prominent system, {Legion}~\cite{Bauer2012}, emphasizes task and data partitioning, enabling automatic parallelism detection and improved locality optimizations. By decoupling computation from hardware mapping, Legion achieves high scalability for complex workloads in distributed and heterogeneous environments. This work utilizes \parsec, which is described in the following section.

\section{The \parsec Runtime System}
\label{sec:bg}

\parsec~\cite{DBLP:journals/pc/BosilcaBDHLD12,parsec-ecp} serves as a generic framework designed for scheduling and managing micro-tasks on distributed, multi-core, and heterogeneous architectures. It integrates a runtime environment, multiple programming models for defining task graphs—applied in dense linear algebra computations~\cite{cao2021evaluating} and irregular workloads such as mixed-precision techniques~\cite{cao2023reducing}, low-rank factorizations~\cite{cao2021leveraging}, and sparse operations~\cite{cao2022framework}—along with auxiliary libraries that ease the transition from legacy implementations. Additionally, it incorporates performance analysis utilities~\cite{cao2019performance} to assist in tuning applications for large-scale heterogeneous platforms.
PaRSEC provides users with a collection of Domain-Specific Languages (DSLs), such as the Parameterized Task Graph (PTG)~\cite{danalis2014ptg}, Template Task Graph (TTG)~\cite{bosilca2020template}, and Dynamic Task Discovery (DTD)~\cite{Hoque:2017:DTD:3148226.3148233}, to enhance flexibility in algorithmic expression. 
Among these, PTG encodes the entire DAG in a compact, parameterized format, enabling optimizations such as collective communication reuse and separation of data distribution from execution. This abstraction aligns with the core methodology of our research on adaptive mixed-precision computations, and we employ PTG in this work.

\section{Algorithm Descriptions}
\label{sec:algo}

\begin{algorithm}[b]
\caption{Tile-centric Mixed-Precision GEMM.}
\label{algo:gemm} 
\SetAlgoLined
\DontPrintSemicolon

\For{$l = 0$ \KwTo $k$}{
    \For{$i = 0$ \KwTo $m$}{
        \For{$j = 0$ \KwTo $n$}{
            $C^*(i, j)=\alpha A^\#(i, l) \times B^\$(l, j)+\beta C^*(i, j)$\;
        }
    }
}
\end{algorithm}

\begin{figure}[t]
    \center
  \begin{subfigure}{0.49\columnwidth}
    \includegraphics[width=1\linewidth]{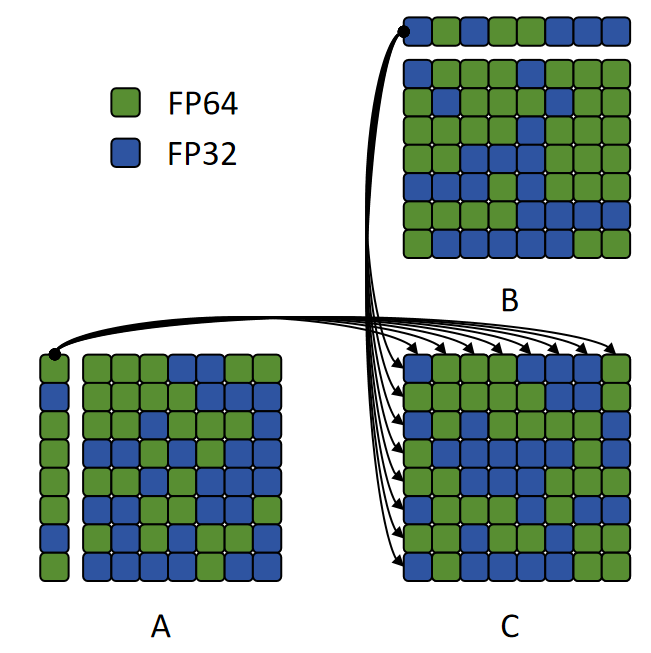}
    \caption{Iteration $k=0$.}
     \label{k=0}
  \end{subfigure}
  \hfill
  \begin{subfigure}{0.49\columnwidth}
    \includegraphics[width=1\linewidth]{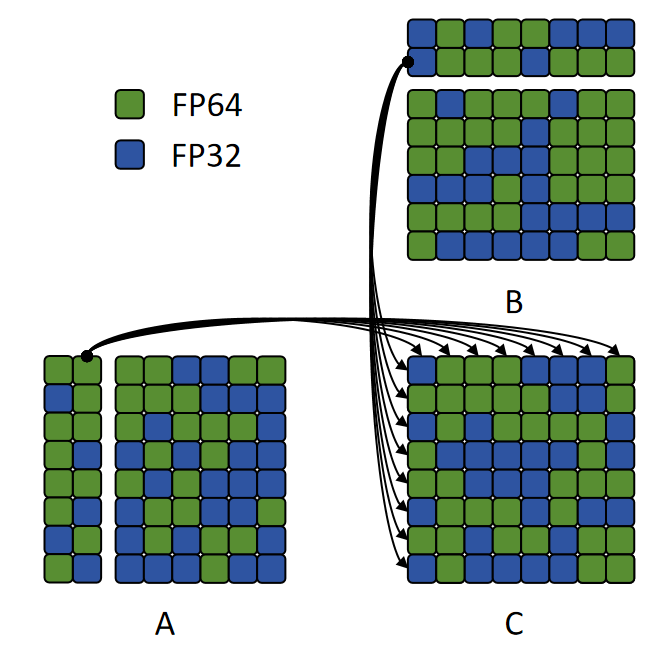}
    \caption{Iteration $k=1$.}
     \label{k=1}
  \end{subfigure}
  \caption{Demonstration of the first two iterations of the mixed-precision GEMM. Arrows are representative dependencies that introduce communications.}
  \label{fig:algo}
  %\vspace{-3ex}
\end{figure}

The General Matrix-Matrix Multiplication (GEMM) operation computes the product of two matrices, optionally scaled and accumulated with a third matrix, and is defined as: $C \leftarrow \alpha AB + \beta C$, where \( A \) and \( B \) are input matrices, \( C \) is the output matrix, and \( \alpha, \beta \) are scalar coefficients. 
Algorithm~\ref{algo:gemm} describes the sequential process of the adaptive mixed-precision GEMM, where the input matrices \( A(i, l)_{m \times k} \), \( B(l, j)_{k \times n} \), and the output matrix \( C(i, j)_{m \times n} \) are partitioned into tiles/blocks of a fixed size. This algorithm follows the SUMMA (Scalable Universal Matrix Multiplication Algorithm) approach~\cite{van1997summa}. 
The outermost loop traverses the reduction dimension \( k \), while the two inner loops iterate over the row index \( i \) and the column index \( j \). At each iteration, the algorithm updates the value of the matrix \( C \). The notations \( \# \), \( \$ \), and \( * \) indicate different data representations or precisions for the operands.

Fig.~\ref{fig:algo} illustrates task execution across the first two iterations as defined in Algorithm~\ref{algo:gemm}. The visualization employs color coding to distinguish numerical precisions: green for double precision (FP64 or DP) and blue for single precision (FP32 or SP). This study focuses on FP64 and FP32, with plans to extend to additional precision formats in future work. Arrows denote representative dependencies arising from data transfer and communication between matrix tiles ($A \rightarrow C$ and $B \rightarrow C$) within a single iteration. Each dependency/communication is associated with a specific datatype. Due to this adaptive tile-centric precision approach, a task may receive data in a precision differing from its operational precision, necessitating potential precision conversion. Various conversion strategies will be explored in subsequent research. This work employs the receiver-side conversion strategy, meaning datatype conversion is handled at the receiving end. That is, the datatype in each data flow is dictated by the precision of the stored data. For example, if a tile in $A$ or $B$ is stored in FP64, the transmitted data remains in double precision.

This algorithm is integrated into the \parsec runtime system, which serves as the underlying framework for execution. By leveraging \parsec, we ensure efficient parallel processing of matrix operations while optimizing the trade-offs among computation, communication, and memory usage, due to the imbalanced workload introduced by the adaptive tile-centric mixed-precision algorithm.

%Due to the differences in precision, the time required for precision conversion and matrix multiplication varies for each tile computation. Therefore, we utilized \parsec for task allocation and scheduling. Specifically, \parsec first decomposes the matrix into tiles and distributes them across different computing nodes following the 2D Block-Cyclic distribution strategy. Each tile multiplication task of A and B is scheduled for execution only when all required input data blocks are ready. Tasks are triggered based on data availability rather than synchronous execution. This dataflow-driven approach eliminates global synchronization, thereby improving parallel efficiency. Additionally, different computational phases within the same tile trigger new tasks, ensuring that computations proceed in the correct order.

%We implement an adaptive mixed-precision GEMM based on the SUMMA algorithm, leveraging \parsec for task scheduling and a 2D Block-Cyclic distribution strategy for efficient parallel execution. The computation follows a dataflow-driven model, ensuring precision consistency across tiles while optimizing data reuse and reducing synchronization overhead in large-scale distributed environments.

\section{Performance Results and Analysis}
\label{sec:perf}

\subsection{Experimental Settings}
The experiments are conducted on three systems.

\begin{itemize}[noitemsep, topsep=0pt]
\item \guyot: It is a GPU-accelerated AMD compute node, which features two EPYC 7742 CPUs, each with 64 cores running at 2.25 GHz, 2,063 GB of main memory, and eight NVIDIA A100-SXM4-80GB GPUs.
%\item Haxane at ICL is a GPU-based Intel compute node. It is equipped with two 8-core Xeon Silver 4309Y CPUs clocked at 2.80 GHz, 63 GB of main memory, and a single NVIDIA H100 PCIe GPU. The deployed CUDA version is 12.1.1.
%\item Leconte at ICL is a GPU-accelerated Intel compute node. It is powered by two Intel Xeon E5-2698 v4 CPUs, each with 20 cores and a base frequency of 2.20 GHz. The compute node is equipped with eight NVIDIA Tesla V100-SXM2-32GB GPUs, each with 32 GB of HBM2 memory. The deployed CUDA version is 12.5.
\item \fugaku: This system is built on ARM architecture and consists of more than 150,000 compute nodes. Each node is powered by a 48-core A64FX CPU, capable of reaching a peak frequency of 2.2 GHz in boost mode and 2.0 GHz in normal mode, and is equipped with 32 GB of HBM2 memory.
\item \frontier: This is a GPU-accelerated supercomputer utilizing AMD hardware, comprising 9,408 compute nodes. Each node integrates a 64-core AMD Optimized 3rd Gen EPYC processor alongside four AMD MI250X GPUs, supported by 512 GB of DDR4 memory.
\end{itemize}

We use the term ``'a'D:'b'S'' to represent the percentage of different precision formats, where D/S denotes double/single precision, and $a$/$b$ represents the percentage of double/single precision, ensuring $a + b = 100$. 
For data distribution, we employ a 2D block cyclic scheme with a process grid of size $P \times Q$, structured to be as square as possible. On \fugaku, we execute in normal mode due to power restrictions and disable the Sector Cache Optimizations (SCO) due to memory conflicts~\cite{cao2022reshaping}. When running on Frontier, we use 4 ranks per node. The optimal tile size is empirically determined, set to 1024 on \fugaku, and 2048 on \guyot and \frontier.

\iffalse
\subsection{Accuracy Evaluation}
To prevent excessive precision loss from causing significant errors in the final result, we designed a verification scheme for the final result matrix. Before executing GEMM, we create an all-FP64-precision copy of matrices $A$, $B$, and $C$. The computation of $C$ is then repeated using the DPLASMA library, and its Frobenius norm is calculated using dplasma\_norm. Next, we compute the difference between the two resulting matrices $C$ and obtain its Frobenfius norm, denoted as R\_norm. Finally, we evaluate the relative difference using $norm\_diff = R\_norm / dplasma\_norm$.
\fi

\subsection{Precision Map of Kernel Execution}

\begin{figure}[t]
    \center
  \begin{subfigure}{0.32\columnwidth}
    \includegraphics[width=1\linewidth]{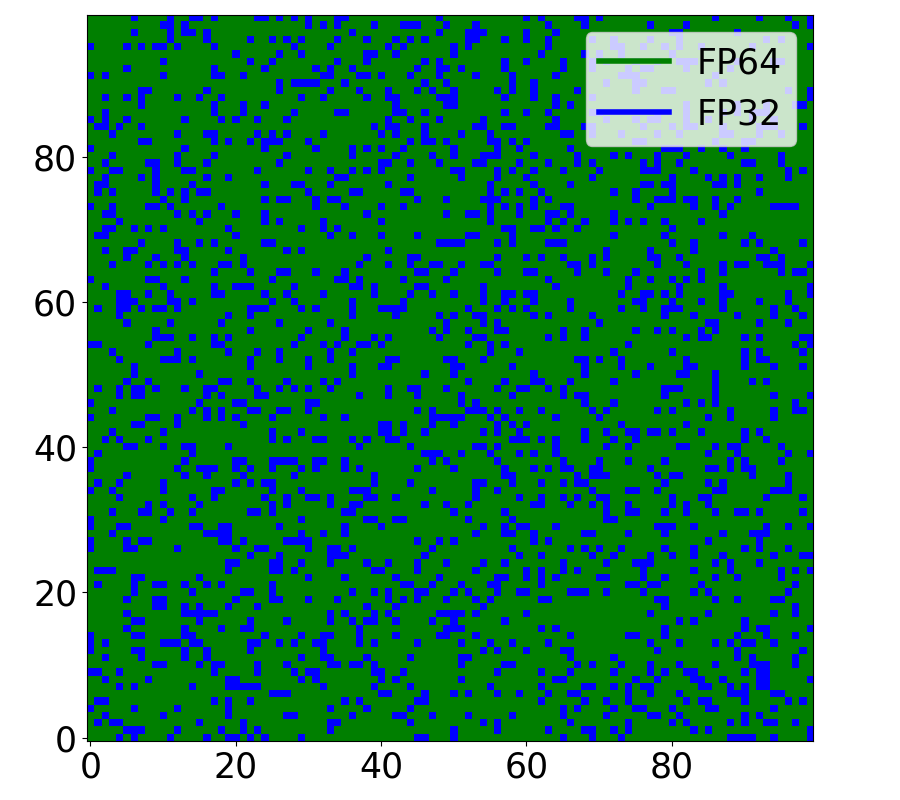}
    \caption{\texttt{80D:20S}.}
     \label{fig:map:dp80}
  \end{subfigure}
  \hfill
  \begin{subfigure}{0.32\columnwidth}
    \includegraphics[width=1\linewidth]{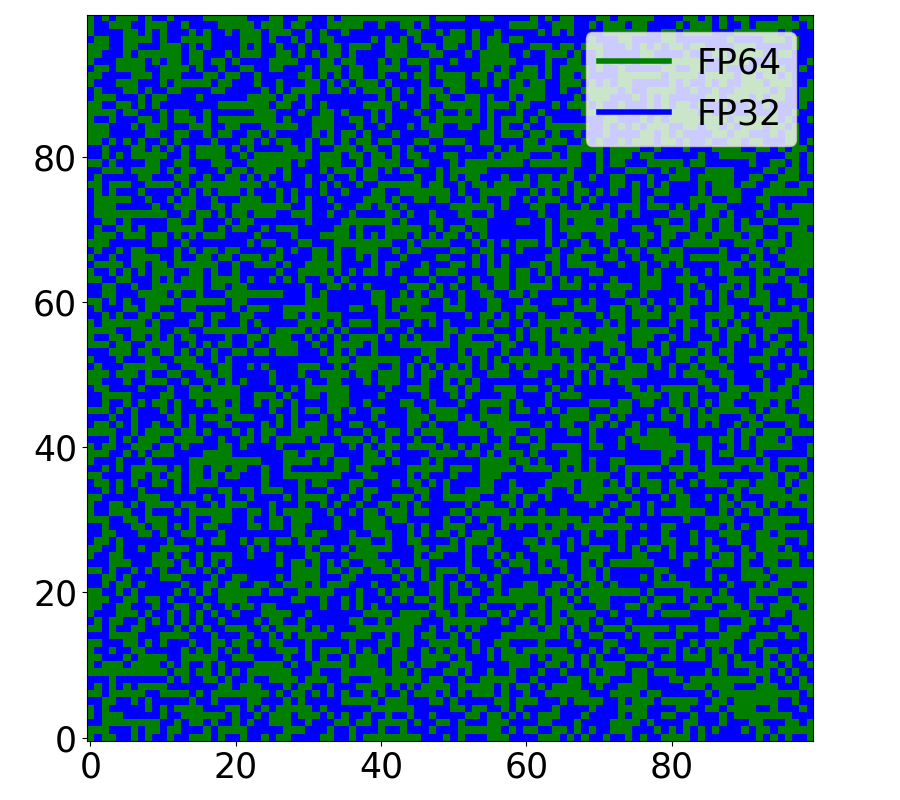}
    \caption{\texttt{50D:50S}.}
     \label{fig:map:dp50}
  \end{subfigure}
    \hfill
    \begin{subfigure}{0.32\columnwidth}
    \includegraphics[width=1\linewidth]{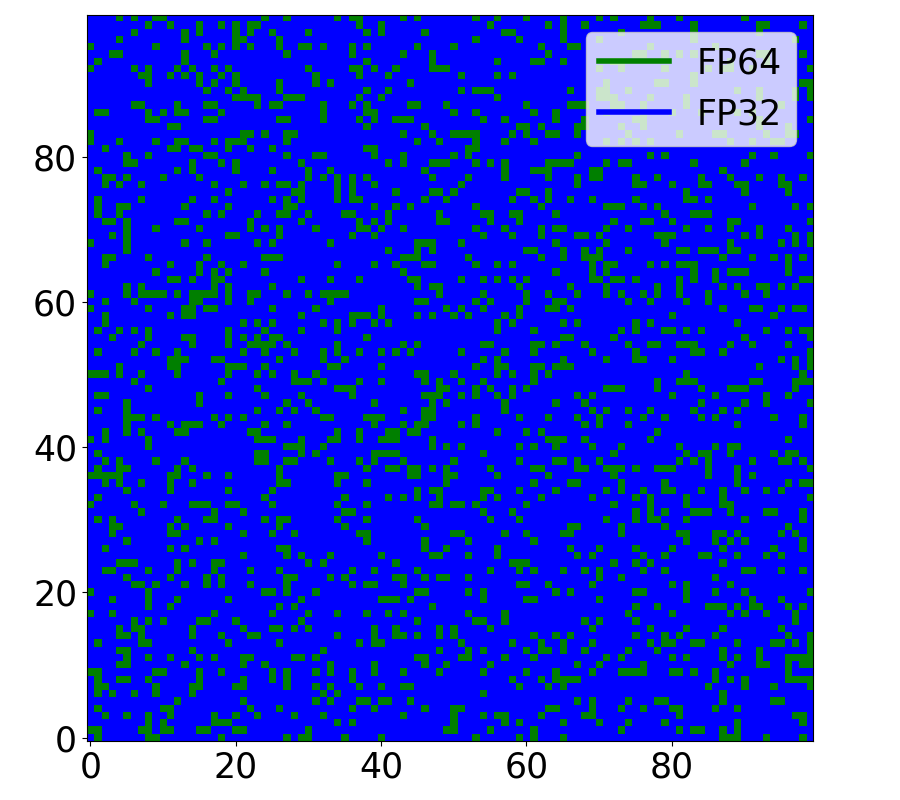}
    \caption{\texttt{20D:80S}.}
     \label{fig:map:dp20}
  \end{subfigure}
  \caption{Kernel precision heatmap for a matrix of size $102,400 \times 102,400$ with a tile size of $1,024 \times 1,024$.}
  \label{fig:map}
  %\vspace{-3ex}
\end{figure}

Fig.~\ref{fig:map} visualizes the precision of numerical kernels executed on each individual tile, as referenced in Fig.~\ref{fig:algo}, by presenting a heatmap under different configurations. Each point represents the precision of a tile in FP64 or FP32, randomly generated in our settings. The three subfigures illustrate different distributions of precision in $A$, $B$, and $C$: (a) \texttt{80D:20S}, where 80\% of the tiles use double precision (FP64) and 20\% use single precision (FP32); (b) \texttt{50D:50S}, where FP64 and FP32 tiles are equally distributed; and (c) \texttt{20D:80S}, where 20\% of the tiles use FP64 and 80\% use FP32. These configurations exhibit varying degrees of mixed-precision computation, which will be employed in the subsequent performance measurements.

%For each tile-partitioned matrix, a 2D integer array, referred to as the precision map, stores precision information. The array’s dimensions correspond to the number of row and column tiles in the matrix, with each integer representing a specific precision (e.g., 0 for FP64 and 1 for FP32). Fig. 2 visualizes tile precision using a heatmap to illustrate the distribution of precision levels across three types of applications.

%To maintain precision information, the precision maps for matrices $A$, $B$, and $C$ are pre-initialized. Each precision map is stored as an unsigned 16-bit integer array, where each entry encodes the precision of a corresponding tile. The initialization of matrix data depends on the precision map. During GEMM execution, the precision maps are referenced to check if precision conversion is necessary for tiles in $A$ and $B$. This approach ensures that the resulting precision of GEMM aligns with the precision specified for the corresponding tile in matrix $C$. 

%For performance evaluation, precision maps for matrices $A$,$ B$, and $C$ are randomly generated based on a predetermined ratio of double precision and single precision. Matrix values are then initialized accordingly, adhering to the precision defined in the maps.

\subsection{Shared Memory Performance}

\begin{figure}[t!]
%\vspace{-4ex}
    \center
  \begin{subfigure}{0.49\columnwidth}
    \includegraphics[width=1\linewidth]{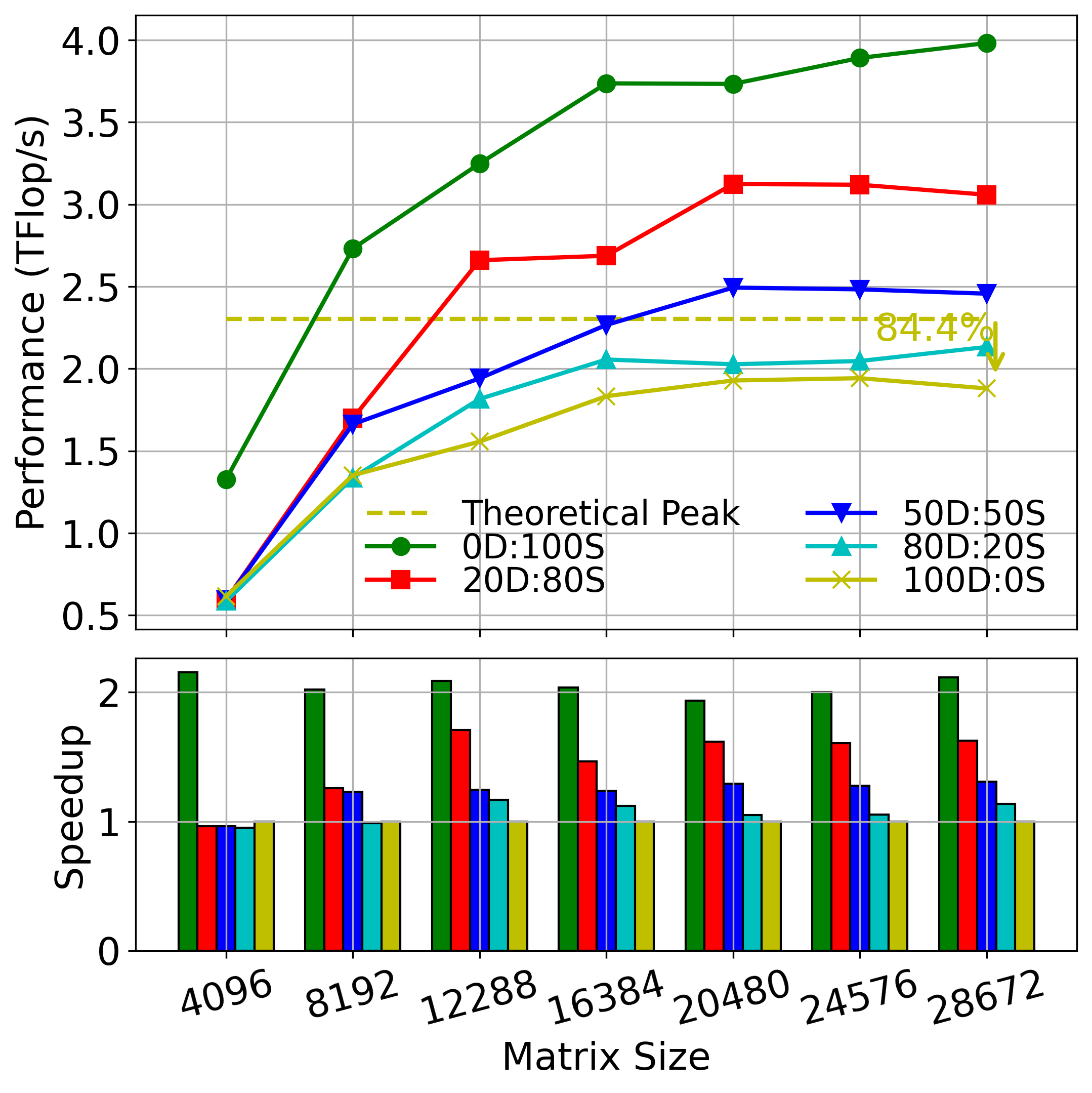}
    \caption{One node on \fugaku.}
     \label{fig:cpu:1node}
  \end{subfigure}
  \hfill
  \begin{subfigure}{0.49\columnwidth}
    \includegraphics[width=1\linewidth]{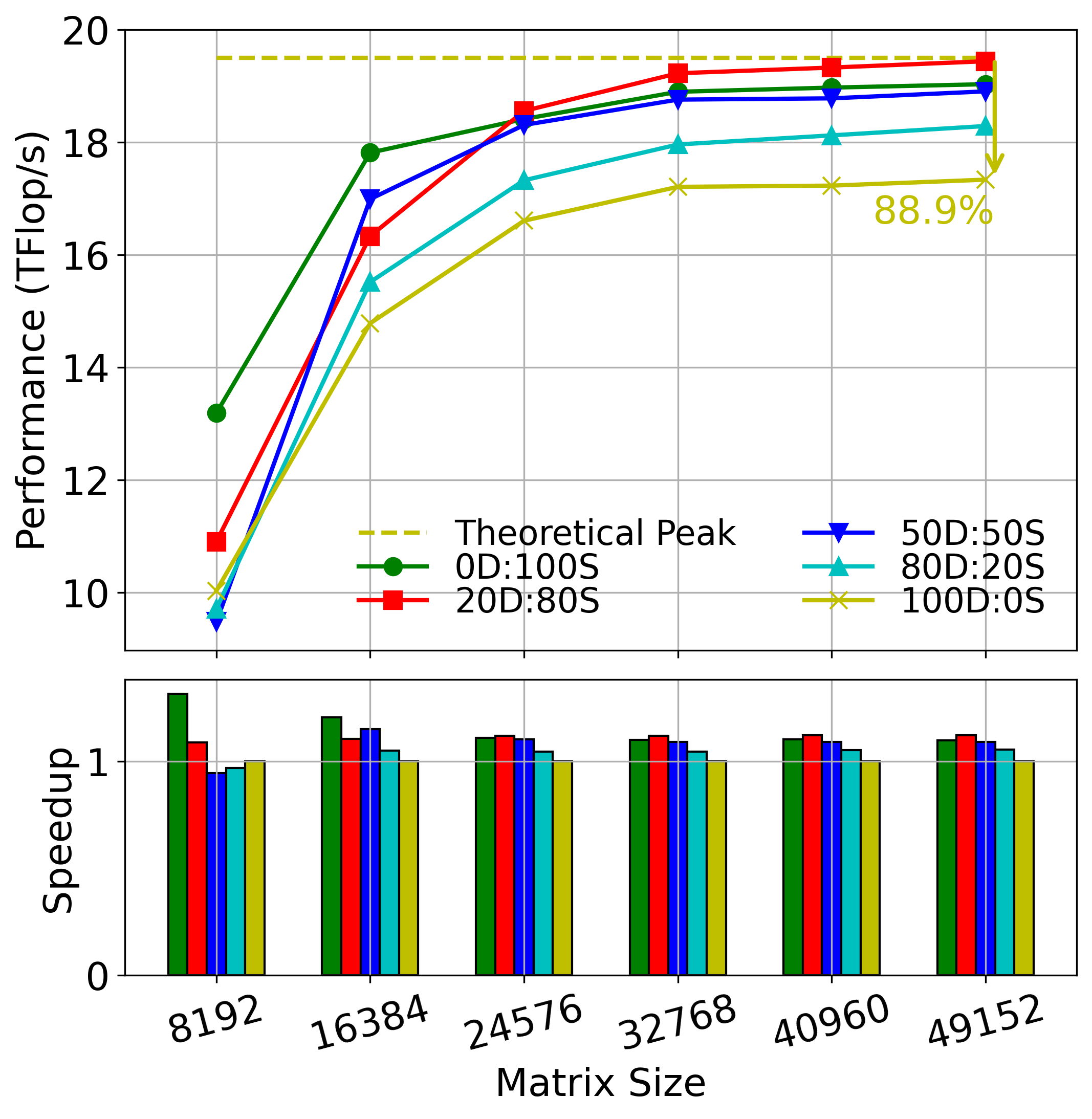}
    \caption{One Nvidia A100 on \guyot.}
     \label{fig:gpu:1A100}
  \end{subfigure}
  \hfill
  \begin{subfigure}{0.49\columnwidth}
    \includegraphics[width=1\linewidth]{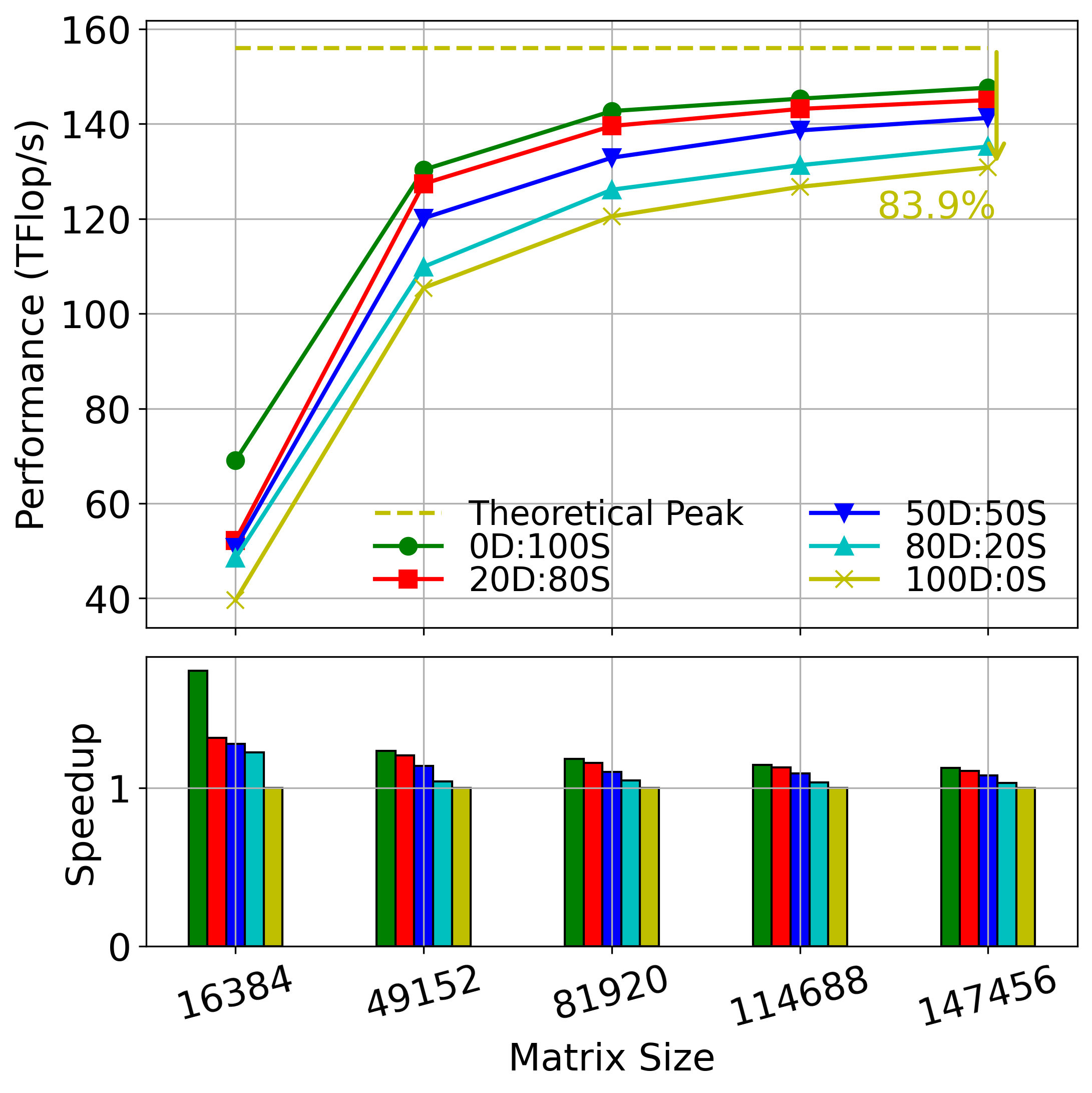}
    \caption{Eight Nvidia A100 on \guyot.}
     \label{fig:gpu:8a100}
  \end{subfigure}
  \hfill
  \begin{subfigure}{0.49\columnwidth}
    \includegraphics[width=1\linewidth]{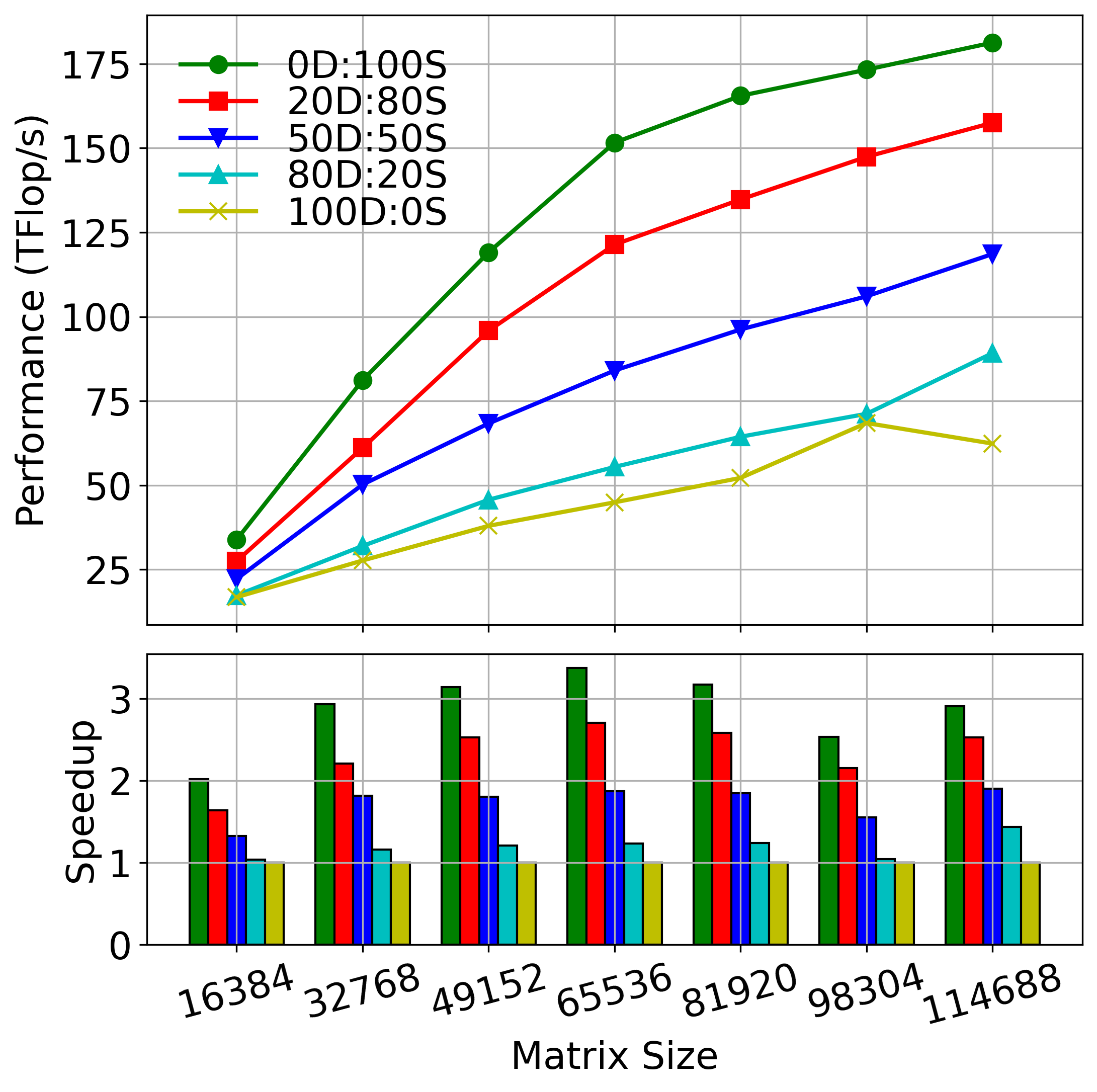}
    \caption{One node on \frontier.}
     \label{fig:gpu:amd_shared}
  \end{subfigure}
  \caption{Shared-memory performance of \ourname.}
  \label{fig:perf:shared}
  %\vspace{-3ex}
\end{figure}

Fig.~\ref{fig:perf:shared} presents the shared-memory performance across different hardware platforms. Each subfigure illustrates performance under various precision settings and the corresponding speedup relative to \texttt{100D:0S}. Different precision configurations exhibit distinct scaling behaviors, with performance generally improving as the proportion of lower precision increases.
(a) Presents performance on a single node of \fugaku. The dashed line represents the practical FP64 peak performance, computed as the GEMM performance per core multiplied by the number of cores. The \texttt{100D:0S} configuration achieves 84.7\% of this practical peak, while \texttt{0D:100S} doubles it.
(b) Depicts results on a single Nvidia A100 GPU on \guyot, where FP64 and FP32 precisions share the same theoretical peak performance. The \texttt{100D:0S} configuration attains up to 88.9\% of the theoretical peak. The decrease in performance with a lower proportion of FP32 may be attributed to increased data movement and/or overheads of datatype conversion.
(c) Demonstrates performance on eight Nvidia A100 GPUs on \guyot, exhibiting an almost linear scaling from the performance of a single GPU (see Fig.~\ref{fig:perf:shared}(b)). The \texttt{100D:0S} configuration reaches 83.9\% of the theoretical peak.
(d) Evaluates performance on a single node of \frontier. The practical peak performance for \texttt{0D:100S} on one node is approximately 280.0 Tflop/s, with an achieved performance of about 181.3 Tflop/s, corresponding to 64.8\% of the practical peak. The \texttt{100D:0S} performance lags a little behind, potentially due to the NIC's proximity to the GPU on \frontier and the ongoing development of GPU-direct communication support in \parsec.

\subsection{Performance Scalability}

\begin{figure}[t!]
    \center
  %
  %begin{subfigure}{0.49\columnwidth}
  %  \includegraphics[width=1\linewidth]{figures/fugaku_4node.png}
  %  \caption{4 nodes on \fugaku.}
  %\end{subfigure}
  %
  %\hfill
  %\begin{subfigure}{0.49\columnwidth}
   % \includegraphics[width=1\linewidth]{figures/fugaku_16node.png}
  %  \caption{16 nodes on \fugaku.}
  %\end{subfigure}
  %
  %\hfill
  \begin{subfigure}{0.468\columnwidth}
    \includegraphics[width=1\linewidth]{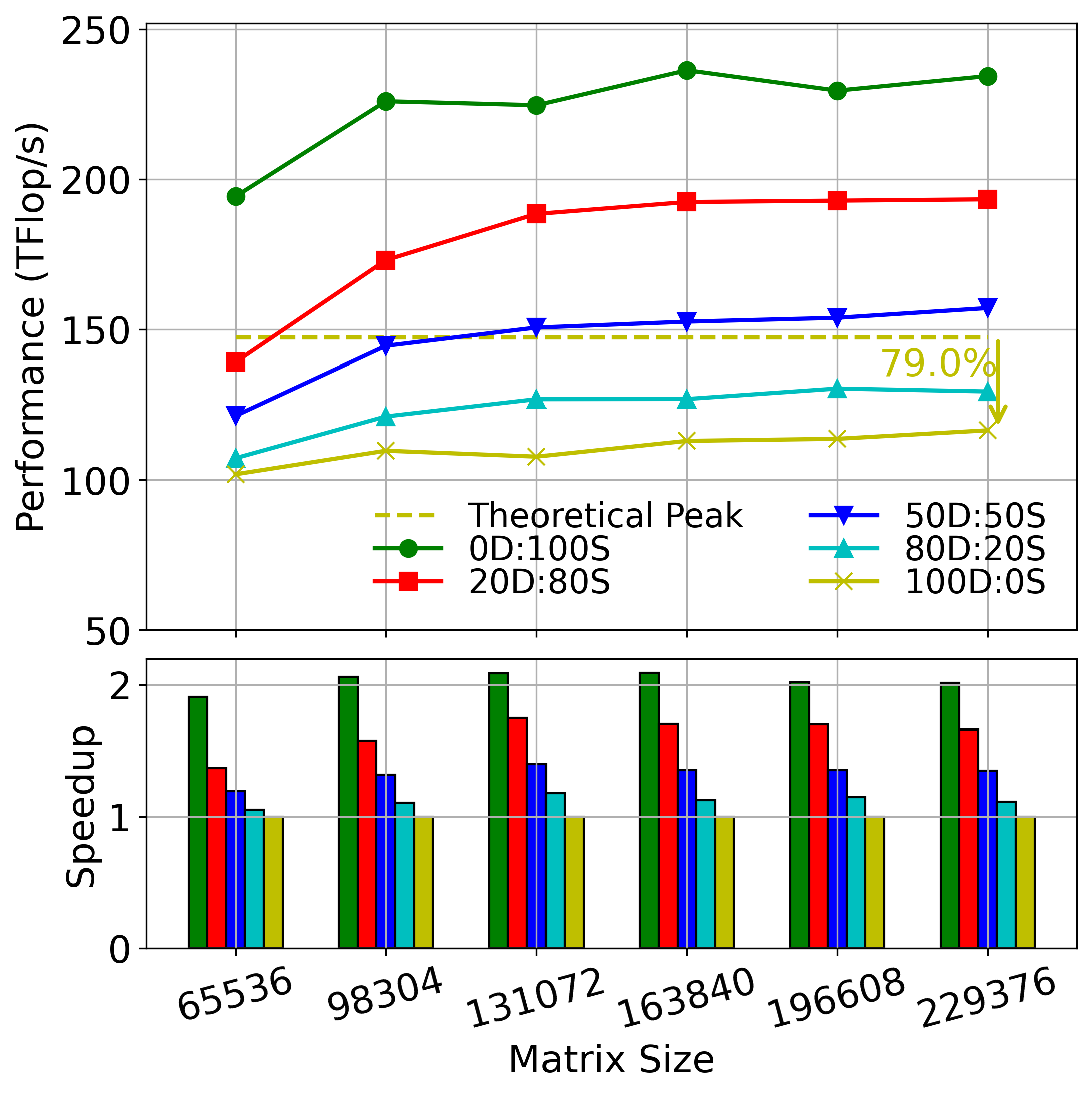}
    \caption{64 nodes on \fugaku.}
  \end{subfigure}
  %
  %\hfill
  %\begin{subfigure}{0.49\columnwidth}
  %  \includegraphics[width=1\linewidth]{figures/frontier_4node.png}
  %  \caption{4 nodes on \frontier.}
  %\end{subfigure}
  %
  %\hfill
  %\begin{subfigure}{0.49\columnwidth}
  %  \includegraphics[width=1\linewidth]{figures/frontier_16node.png}
  %  \caption{16 nodes on \frontier.}
  %\end{subfigure}
  %
  \hfill
  \begin{subfigure}{0.49\columnwidth}
    \includegraphics[width=1\linewidth]{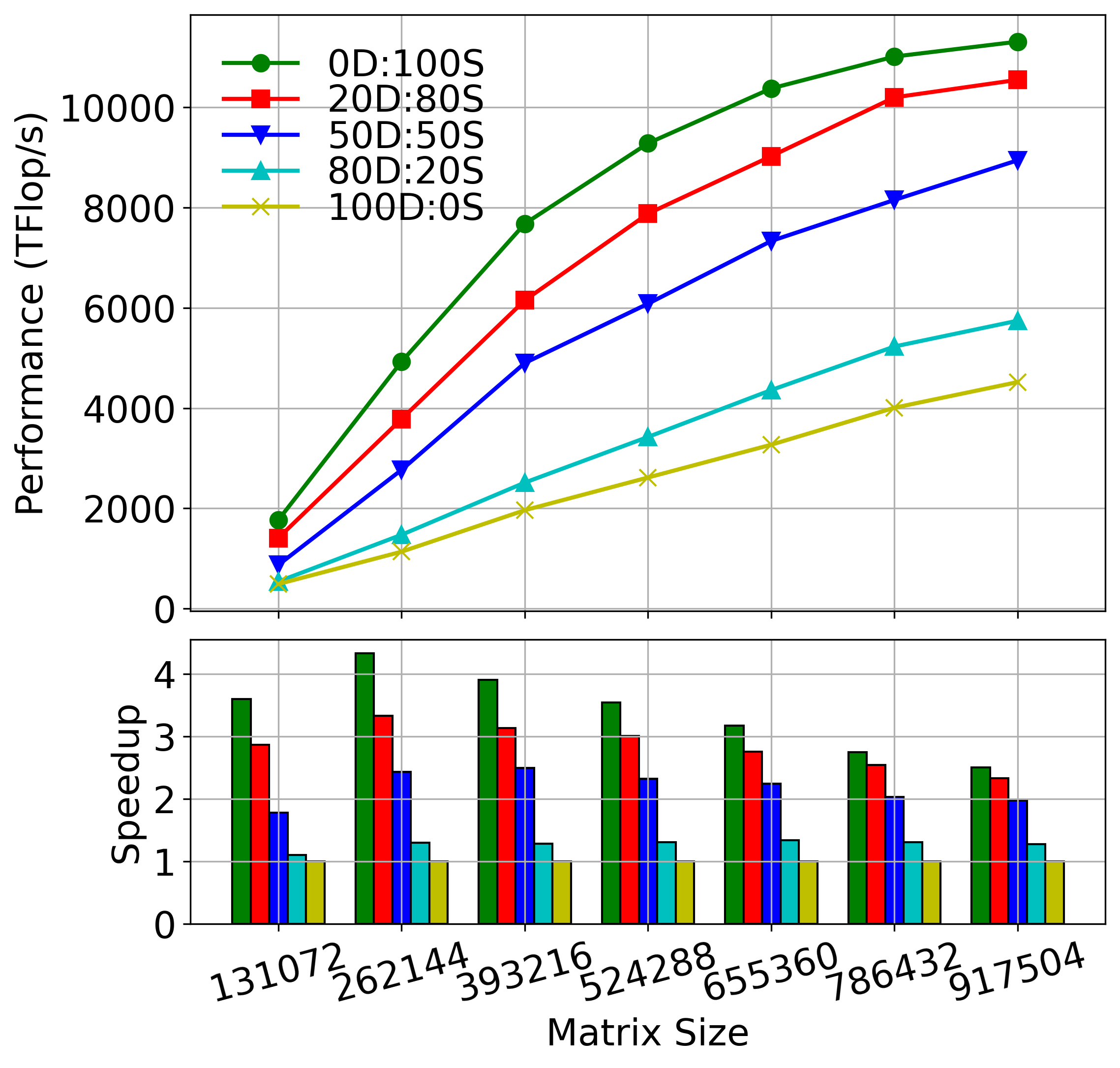}
    \caption{64 nodes on \frontier.}
  \end{subfigure}
  \caption{Distributed-memory performance of \ourname.}
  \label{fig:perf:distributed}
  %\vspace{-3ex}
\end{figure}

%Fig.~\ref{fig:pef:distributed} presents the performance results on \fugaku and \frontier with 64 nodes.

%On \fugaku, the performance of full FP64 matrices reaches approximately 80\% of the theoretical peak. Performance improves as the proportion of FP32 tiles increases, with full FP32 matrices achieving about 2× speedup compared to full FP64 matrices, indicating that performance is primarily constrained by communication overhead. While communication overhead is high for small-scale computations, performance remains stable even when scaling up to 64 nodes.

%On \frontier, although the performance of full FP64 matrices reaches less than 40\% of the theoretical peak, increasing the proportion of FP32 tiles leads to a more significant speedup, achieving over 2.50× improvement. Additionally, for full FP32 matrices, performance approaches the theoretical peak. Furthermore, as the computational scale increases, performance continues to scale efficiently, demonstrating strong weak scalability.

Fig.~\ref{fig:perf:distributed} presents the distributed-memory performance on 64 nodes of \fugaku and \frontier, further confirming the trends observed in shared-memory performance, as shown in Fig.~\ref{fig:perf:shared}: the achieved performance approaches the theoretical/practical peak, and different precision configurations exhibit distinct scaling behaviors, with performance generally improving as the proportion of lower precision increases.
Moreover, \ourname attains nearly linear scaling from shared-memory performance to 64 nodes on both \frontier and \fugaku. For instance, in the \texttt{0D:100S} configuration, the one-node performance on \fugaku is approximately 4.0 Tflop/s, while the performance on 64 nodes reaches 242.2 Tflop/s with a parallel efficiency of 94.6\%. On \frontier, the one-node performance is about 181.3 Tflop/s, increasing to 11,310.3 Tflop/s on 64 nodes with a parallel efficiency of 97.5\%.
All in all, these results underscore the significance of the proposed tile-centric mixed-precision framework in achieving optimal computational efficiency in both shared-memory and distributed-memory environments across various architectures.

\section{Conclusion and Future Work}
\label{sec:conclusion}

This work introduces \ourname, an adaptive tile-centric mixed-precision GEMM framework. Built on \parsec, the proposed approach efficiently manages heterogeneous tasks arising from varying precision formats, ensuring high-performance execution across both shared- and distributed-memory environments. \ourname demonstrates strong efficiency and scalability on homogeneous CPU setups and heterogeneous GPU systems, making it highly adaptable to diverse hardware architectures.
Going forward, we plan to extend the framework by incorporating additional precision formats and developing trustworthy precision selection strategies for both lossless and lossy approaches. We also aim to investigate different datatype conversion strategies. Furthermore, we seek to integrate this mixed-precision framework with other runtime systems and deploy it in real-world HPC and AI applications.

\section*{Acknowledgments}

This research was supported by internal awards from Saint Louis University (Grant-0001651 and PROJ-000498) and the U.S. National Science Foundation (Award OAC-2451577). For computer time, this research used the Lonestar6 cluster from Texas Advanced Computing Center, the compute node at Innovative Computing Laboratory of the University of Tennessee, Knoxville, the Fugaku supercomputer at RIKEN, and Frontier supercomputer at Oak Ridge National Laboratory.

\bibliographystyle{unsrt}
\bibliography{ref}

\begin{thebibliography}{10}

\bibitem{ichimura2018fast}
Tsuyoshi Ichimura, Kohei Fujita, Takuma Yamaguchi, Akira Naruse, Jack~C Wells, Thomas~C Schulthess, Tjerk~P Straatsma, Christopher~J Zimmer, Maxime Martinasso, Kengo Nakajima, et~al.
\newblock A fast scalable implicit solver for nonlinear time-evolution earthquake city problem on low-ordered unstructured finite elements with artificial intelligence and transprecision computing.
\newblock In {\em SC18: International Conference for High Performance Computing, Networking, Storage and Analysis}, pages 627--637. IEEE, 2018.

\bibitem{abdulah2024boosting}
Sameh Abdulah, Allison~H. Baker, George Bosilca, Qinglei Cao, Stefano Castruccio, Marc~G. Genton, David~E. Keyes, Zubair Khalid, Hatem Ltaief, Georgiy L.~Stenchikov Yan~Song, and Ying Sun.
\newblock Boosting earth system model outputs and saving petabytes in their storage using exascale climate emulators.
\newblock {\em ACM Gordon Bell Prize for Climate Modelling Finalist}, 2024.

\bibitem{vaswani2017attention}
A~Vaswani.
\newblock Attention is all you need.
\newblock {\em Advances in Neural Information Processing Systems}, 2017.

\bibitem{top500}
"Hans Meuer, Erich Strohmaier, Jack Dongarra, and Horst Simon".
\newblock {The Top500 List}.
\newblock June 2024.
\newblock {\url{http://www.top500.org}}.

\bibitem{abdulah2021accelerating}
Sameh Abdulah, Qinglei Cao, Yu~Pei, George Bosilca, Jack Dongarra, Marc~G Genton, David~E Keyes, Hatem Ltaief, and Ying Sun.
\newblock Accelerating geostatistical modeling and prediction with mixed-precision computations: A high-productivity approach with parsec.
\newblock {\em IEEE Transactions on Parallel and Distributed Systems}, 33(4):964--976, 2021.

\bibitem{cao2023reducing}
Qinglei Cao, Sameh Abdulah, Hatem Ltaief, Marc~G Genton, David Keyes, and George Bosilca.
\newblock Reducing data motion and energy consumption of geospatial modeling applications using automated precision conversion.
\newblock In {\em 2023 IEEE International Conference on Cluster Computing (CLUSTER)}, pages 330--342. IEEE, 2023.

\bibitem{haidar2018design}
Azzam Haidar, Ahmad Abdelfattah, Mawussi Zounon, Panruo Wu, Srikara Pranesh, Stanimire Tomov, and Jack Dongarra.
\newblock The design of fast and energy-efficient linear solvers: On the potential of half-precision arithmetic and iterative refinement techniques.
\newblock In {\em Computational Science--ICCS 2018: 18th International Conference, Wuxi, China, June 11--13, 2018, Proceedings, Part I}, pages 586--600. Springer, 2018.

\bibitem{haidar2020mixed}
Azzam Haidar, Harun Bayraktar, Stanimire Tomov, Jack Dongarra, and Nicholas~J Higham.
\newblock Mixed-precision iterative refinement using tensor cores on gpus to accelerate solution of linear systems.
\newblock {\em Proceedings of the Royal Society A}, 476(2243):20200110, 2020.

\bibitem{netti2023mixed}
Alessio Netti, Yang Peng, Patrik Omland, Michael Paulitsch, Jorge Parra, Gustavo Espinosa, Udit Agarwal, Abraham Chan, and Karthik Pattabiraman.
\newblock Mixed precision support in hpc applications: What about reliability?
\newblock {\em Journal of Parallel and Distributed Computing}, 181:104746, 2023.

\bibitem{nandakumar2020mixed}
SR~Nandakumar, Manuel Le~Gallo, Christophe Piveteau, Vinay Joshi, Giovanni Mariani, Irem Boybat, Geethan Karunaratne, Riduan Khaddam-Aljameh, Urs Egger, Anastasios Petropoulos, et~al.
\newblock Mixed-precision deep learning based on computational memory.
\newblock {\em Frontiers in neuroscience}, 14:406, 2020.

\bibitem{walden2019mixed}
Aaron Walden, Eric Nielsen, Boris Diskin, and Mohammad Zubair.
\newblock A mixed precision multicolor point-implicit solver for unstructured grids on gpus.
\newblock In {\em 2019 IEEE/ACM 9th Workshop on Irregular Applications: Architectures and Algorithms (IA3)}, pages 23--30. IEEE, 2019.

\bibitem{brogi2024floating}
Federico Brogi, Simone Bn{\`a}, Gabriele Boga, Giorgio Amati, T~Esposti Ongaro, and Matteo Cerminara.
\newblock On floating point precision in computational fluid dynamics using openfoam.
\newblock {\em Future Generation Computer Systems}, 152:1--16, 2024.

\bibitem{doucet2019mixed}
Nicolas Doucet, Hatem Ltaief, Damien Gratadour, and David Keyes.
\newblock Mixed-precision tomographic reconstructor computations on hardware accelerators.
\newblock In {\em 2019 IEEE/ACM 9th Workshop on Irregular Applications: Architectures and Algorithms (IA3)}, pages 31--38. IEEE, 2019.

\bibitem{chen2024mixed}
Siyuan Chen, Yi~Zhang, Yiming Wang, Zhuang Liu, Xiaohan Li, and Wei Xue.
\newblock Mixed-precision computing in the grist dynamical core for weather and climate modelling.
\newblock {\em Geoscientific Model Development Discussions}, 2024:1--28, 2024.

\bibitem{jia2020pushing}
Weile Jia, Han Wang, Mohan Chen, Denghui Lu, Lin Lin, Roberto Car, E~Weinan, and Linfeng Zhang.
\newblock Pushing the limit of molecular dynamics with ab initio accuracy to 100 million atoms with machine learning.
\newblock In {\em SC20: International conference for high performance computing, networking, storage and analysis}, pages 1--14. IEEE, 2020.

\bibitem{boku2017mixed}
Taisuke Boku, Ken-Ichi Ishikawa, Yoshinobu Kuramashi, and Lawrence Meadows.
\newblock Mixed precision solver scalable to 16000 mpi processes for lattice quantum chromodynamics simulations on the oakforest-pacs system.
\newblock In {\em 2017 Fifth International Symposium on Computing and Networking (CANDAR)}, pages 362--368. IEEE, 2017.

\bibitem{haidar2018harnessing}
Azzam Haidar, Stanimire Tomov, Jack Dongarra, and Nicholas~J Higham.
\newblock Harnessing gpu tensor cores for fast fp16 arithmetic to speed up mixed-precision iterative refinement solvers.
\newblock In {\em SC18: International Conference for High Performance Computing, Networking, Storage and Analysis}, pages 603--613. IEEE, 2018.

\bibitem{cao2022reshaping}
Qinglei Cao, Sameh Abdulah, Rabab Alomairy, Yu~Pei, Pratik Nag, George Bosilca, Jack Dongarra, Marc~G Genton, David~E Keyes, Hatem Ltaief, and Ying Sun.
\newblock Reshaping geostatistical modeling and prediction for extreme-scale environmental applications.
\newblock In {\em SC22: International Conference for High Performance Computing, Networking, Storage and Analysis (ACM Gordon Bell Prize Finalist)}, pages 1--12. IEEE, 2022.

\bibitem{hatem2024toward}
Hatem Ltaief, Rabab Alomairy, Qinglei Cao, Jie Ren, Lotfi Slim, Thorsten Kurth, Benedikt Dorschner, Salim Bougouffa, Rached Abdelkhalek, and David~E. Keyes.
\newblock Toward capturing genetic epistasis from multivariate genome-wide association studies using mixed-precision kernel ridge regression.
\newblock {\em ACM Gordon Bell Prize Finalist}, 2024.

\bibitem{OpenMP}
OpenMP.
\newblock {OpenMP 4.5 Complete Specifications}, 2015.

\bibitem{ompss}
A.~Duran, R.~Ferrer, E.~Ayguade, R.~M. Badia, and J.~Labarta.
\newblock A proposal to extend the {OpenMP} tasking model with dependent tasks.
\newblock {\em Intl. Journal of Parallel Programming}, 37(3):292--305, 2009.

\bibitem{lordan2014servicess}
Francesc Lordan, Enric Tejedor, Jorge Ejarque, Roger Rafanell, Javier Alvarez, Fabrizio Marozzo, Daniele Lezzi, Ra{\"u}l Sirvent, Domenico Talia, and Rosa~M Badia.
\newblock {Servicess: An Interoperable Programming Framework for the Cloud}.
\newblock {\em Journal of Grid Computing}, 2014.

\bibitem{starpu}
C.~{Augonnet}, S.~{Thibault}, R.~{Namyst}, and P.~{Wacrenier}.
\newblock {StarPU}: A unified platform for task scheduling on heterogeneous multicore architectures.
\newblock {\em Concurrency Computat. Pract. Exper.}, 23:187--198, 2011.

\bibitem{Heller2013a}
T.~Heller, H.~Kaiser, and K.~Iglberger.
\newblock {Application of the ParalleX execution model to stencil-based problems}.
\newblock {\em Computer Science - Research and Development}, 28(2-3):253--261, 2013.

\bibitem{Bauer2012}
Michael Bauer, Sean Treichler, Elliott Slaughter, and Alex Aiken.
\newblock {Legion}: Expressing locality and independence with logical regions.
\newblock In {\em International Conference for High Performance Computing, Networking, Storage and Analysis, SC}, pages 1--11. IEEE, 2012.

\bibitem{DBLP:journals/pc/BosilcaBDHLD12}
George Bosilca, Aur{\'{e}}lien Bouteiller, Anthony Danalis, Thomas H{\'{e}}rault, Pierre Lemarinier, and Jack~J. Dongarra.
\newblock {DAGuE}: {A} generic distributed {DAG} engine for high performance computing.
\newblock {\em Parallel Comput.}, 38(1-2):37--51, 2012.

\bibitem{parsec-ecp}
Aurelien Bouteiller, Thomas Herault, Qinglei Cao, Joseph Schuchart, and George Bosilca.
\newblock {PaRSEC}: Scalability, flexibility, and hybrid architecture support for task-based applications in {ECP}.
\newblock {\em International Journal of High Performance Computing Applications}, 2024.

\bibitem{cao2021evaluating}
Qinglei Cao, George Bosilca, Nuria Losada, Wei Wu, Dong Zhong, and Jack Dongarra.
\newblock Evaluating data redistribution in parsec.
\newblock {\em IEEE Transactions on Parallel and Distributed Systems}, 33(8):1856--1872, 2021.

\bibitem{cao2021leveraging}
Qinglei Cao, Yu~Pei, Kadir Akbudak, George Bosilca, Hatem Ltaief, David Keyes, and Jack Dongarra.
\newblock Leveraging parsec runtime support to tackle challenging 3d data-sparse matrix problems.
\newblock In {\em 2021 IEEE International Parallel and Distributed Processing Symposium (IPDPS)}, pages 79--89. IEEE, 2021.

\bibitem{cao2022framework}
Qinglei Cao, Rabab Alomairy, Yu~Pei, George Bosilca, Hatem Ltaief, David Keyes, and Jack Dongarra.
\newblock A framework to exploit data sparsity in tile low-rank cholesky factorization.
\newblock In {\em 2022 IEEE International Parallel and Distributed Processing Symposium (IPDPS)}, pages 414--424. IEEE, 2022.

\bibitem{cao2019performance}
Qinglei Cao, Yu~Pei, Thomas Herault, Kadir Akbudak, Aleksandr Mikhalev, George Bosilca, Hatem Ltaief, David Keyes, and Jack Dongarra.
\newblock Performance analysis of tile low-rank cholesky factorization using parsec instrumentation tools.
\newblock In {\em 2019 IEEE/ACM International Workshop on Programming and Performance Visualization Tools (ProTools)}, pages 25--32. IEEE, 2019.

\bibitem{danalis2014ptg}
Anthony Danalis, George Bosilca, Aurelien Bouteiller, Thomas Herault, and Jack Dongarra.
\newblock {PTG: an Abstraction for Unhindered Parallelism}.
\newblock In {\em 2014 Fourth International Workshop on Domain-Specific Languages and High-Level Frameworks for High Performance Computing}, pages 21--30. IEEE, 2014.

\bibitem{bosilca2020template}
George Bosilca, Robert~J Harrison, Thomas Herault, Mohammad~Mahdi Javanmard, P~Nookala, and Edward~F Valeev.
\newblock {The Template Task Graph (TTG)-an Emerging Practical Dataflow Programming Paradigm for Scientific Simulation at Extreme Scale}.
\newblock In {\em IEEE/ACM 5th International Workshop on Extreme Scale Programming Models and Middleware (ESPM2)}. IEEE, 2020.

\bibitem{Hoque:2017:DTD:3148226.3148233}
R.~Hoque, T.~Herault, G.~Bosilca, and J.~Dongarra.
\newblock {Dynamic Task Discovery in PaRSEC: A Data-flow Task-based Runtime}.
\newblock In {\em Proceedings of the 8th Workshop on Latest Advances in Scalable Algorithms for Large-Scale Systems}, ScalA '17, 2017.

\bibitem{van1997summa}
Robert~A Van De~Geijn and Jerrell Watts.
\newblock Summa: Scalable universal matrix multiplication algorithm.
\newblock {\em Concurrency: Practice and Experience}, 9(4):255--274, 1997.

\end{thebibliography}

\end{document}